\documentclass[preprint,12pt]{elsarticle}
\usepackage{xcolor}

\usepackage{amssymb}
\usepackage{amsmath}

\journal{Optics Communications}

\begin{document}

\begin{frontmatter}



\title{Dynamics of Light Localization via Coherent Control: The Interplay of Transmission, Absorption and Disorder in Photonic Crystals}


\author[a]{Nancy Ghangas \corref{cor1}}
\cortext[cor1]{Corresponding author. Email: nancy.19phz0006@iitrpr.ac.in}
\author[a]{Shubhrangshu Dasgupta} 
\affiliation[a]{organization={{Department of Physics}},
            addressline={Indian Institute of Technology Ropar}, 
            city={Rupnagar},
            postcode={140001}, 
            state={Punjab},
            country={India}}
\author[b]{Ghanasyam Remesh}
\author[b]{Venu Gopal Achanta}
\affiliation[b]{organization={{Department of Condensed Matter Physics and Material Science}},
            addressline={Tata Institute of Fundamental Research}, 
            city={Mumbai},
            postcode={400006}, 
            state={Maharashtra},
            country={India}}
\begin{abstract}
This study investigates the interplay between structural disorder, absorption, and Lyapunov exponent dynamics to exploit localization phenomena in photonic crystals with engineered defect layers. We generate disorder by introducing random refractive index variations in one of the bilayers, while the application of a control field to $\Lambda$-type atoms within a central defect layer enables dynamic tuning of the crystal’s effective refractive index. We have employed traditional transfer matrix method to demonstrate transmission, Lyapunov exponents and absorption in the crystal. Through coherent control, we dynamically tune absorption, revealing sharp contrasts in band gap and band edge regions. while Lyapunov exponents, quantifying localization lengths, exhibit a consistent scaling across both band gap and band edge frequencies, and this behavior remains robust even in the presence of disorder. Hence, distinct localization mechanisms emerge at bandgap and band-edge frequencies. Bandgap localization arises from optical mode confinement and resonant alignment of atomic transitions with the probe field while band edge localization stems from a synergy of loss-difference-induced trapping and Anderson-like disorder effects. Notably, while disorder weakens confinement localization in the band gap, it actually strengthens localization at the band edges. These results deepen the understanding of light-matter coupling in disordered photonic systems and provide a framework for designing reconfigurable optical devices with tailored localization properties.
\end{abstract}


\begin{highlights}
\item Active tuning of absorption, Lyapunov exponents (quantifying confinement stability), and localization regimes via coherent interactions, combining refractive index randomness with $\Lambda$-atom doping in a defect layer.
\item Contrast in band gap and band edge localization mechanisms, absorption enhances confinement (disrupted by disorder) in band gap while disorder boosts light trapping in band edges.
\end{highlights}


\end{frontmatter}


\section{Introduction}\label{sec1}

The localization of light in Photonic Crystals (PCs) arises when structural irregularities trap light waves, disrupting their normal propagation and reducing their spread over large distances \cite{PhysRev.109.1492}. Yablonovitch's seminal work revealed that three-dimensional periodic dielectric structures with photonic band gaps can suppress spontaneous emission by preventing radiative transitions within the gap \cite{PhysRevLett.58.2059}. Further, it has also been proposed that carefully designed disorder in 3D dielectric superlattices could achieve controlled photon localization in lossless systems, leveraging deliberate randomness rather than energy dissipation, to confine light effectively \cite{PhysRevLett.58.2059}. PCs enable advanced applications in optical communications, sensing, and energy control by precisely engineering light–matter interactions.  By controlling light in engineered photonic crystals and metastructures, advanced photonic devices to achieve precise privacy filtering \cite{Pei_2024}, asymmetric absorption-transmission \cite{LEI2024101810}, and graphene-based electromagnetic detection via second harmonic generation \cite{10382952} have been developed. Although the critical disorder level to achieve such control and the role of interactions in such processes remained elusive for decades, modern experiments using ultracold atoms in disordered optical potentials add insights on Anderson localization \cite{10.1063/1.3206092}.
Various studies on PCs have combined experimental and computational approaches to analyze how controlled structural disorders,  both correlated and uncorrelated, affect their optical properties \cite{Liew_2010,PhysRevA.84.023813,PhysRevB.95.224202,PhysRevA.100.033804,NAYAK201725,PhysRevA.103.023503,GHANGAS2024115703}. A few reports have also examined linear and nonlinear wave propagation, slow light transport, and angle-dependent effects in disordered 1D waveguide arrays, demonstrating realizations of the Anderson localization \cite{PhysRevLett.100.013906,PhysRevLett.103.063903, Petrov:09,SARABANDI2020126263,PhysRevB.41.8047}.
Despite ongoing experimental challenges in fabricating such disordered systems, recent advances have introduced innovative techniques such as stochastic layer stacking with controlled randomness, defect-layer doping with stimuli-responsive materials \cite{doi:10.1021/acsnano.1c11505}, flexible integration of polymeric–nanoparticle multilayers on stretchable substrates \cite{KRIEGEL2022111859,BELLINGERI2015235}, and twist-induced disorder in moiré PCs \cite{Alnasser_2024}. These methods not only mimic the adaptive sensing capabilities of human skin but also enhance optical performance for wearable devices, LED light extraction, and sensitive biosensing (e.g., blood detection and tumor screening) \cite{Chen01012017,Pradhan:23,photonics10111199}. A thorough review of these fabrication, design, and measurement challenges in such advanced photonic structures have been discussed \cite{ijms242115790}.
 
Optical Mode Confinement governs the behavior of photons in such structured dielectric materials. It involves manipulating light propagation through periodic refractive index variations and engineered defects, creating photonic bandgaps that trap or guide light in localized modes as a result of interference. These modes confine light to sub-wavelength volumes, enhancing light-matter interactions via Purcell Effect which is the spontaneous emission rate that quantifies absorption enhancement in a defect mode. It scales with the local photonic density of states (DOS) in frequency spectrum \cite{PhysRevLett.58.2059}. 
The authors report that disorder in PCs substantially modifies the Purcell effect where moderate disorder increases spontaneous emission at the band gap edges, while stronger disorder generates high-Q states that enhance emission within the gap, leading to mirrorless lasing at the band edge and superlinear emission behavior in synthetic opals \cite{Morozov_2018}.
In contrast to optical confinement, quantum confinement restricts electrons and holes in nanoscale structures (e.g., quantum dots), forcing energy levels to become discrete and altering the material's optical and electronic properties. This confinement compresses multiple nearby bulk transitions into a single, intense atomic transition, where localized modes have well-defined energy and position but increased momentum uncertainty \cite{doi:10.1126/science.271.5251.933}. Optical confinement occurs at microscale (hundreds of nanometers to micrometers), aligned with the wavelength of light while quantum confinement happens at nano size confinement ($1-100nm$). However, both optical and quantum confinement modify the system's properties, including emission and absorption.
A recent study demonstrates that doping PCs with trivalent Europium ions enables non-Hermitian alignment of energy levels, allowing precise control over constructive and destructive optical confinement by tuning laser parameters, polarization, and decay rates\cite{D4TC03660C}. 

In disordered PCs, Lyapunov exponents (LEs) quantify how quickly the intensity of light decays as it propagates through a disordered medium due to interference. They are associated with the inverse of the localization length and are critical to understanding wave transport phenomena. In general, LEs are numbers that measure how quickly nearby states in a dynamical system diverge or converge over time. In PCs, the localization length tends to be greater in the propagating regions and lower in the gap regions, leading to the formation of wave functions that are confined to limited spatial confines. It has been shown with LE dynamics that when disorder surpasses a critical threshold, states from pass bands and band edges exhibit single parameter scaling behavior but states from mid bandgap deviate from this behavior \cite{PhysRevLett.81.5390}. Subsequent studies have predicted distinct localization regimes in such disordered PCs \cite{PhysRevB.60.1555}. Researchers have also exploited the effects of disorder in refractive indices in 1D PCs to quantitatively tailor the localization and delocalization of light within the photonic bandgap and their associated passbands \cite{Yuan:19}. Although, the applications of LE is not without constraints and their limitations can impact the interpretation and reliability of results in certain scenarios \cite{Comtet_2013}.
In such photonic systems, it has been reported that wave decay can occur due to various phenomena, including localization, inelastic scattering, and out-of-plane losses \cite{PhysRevA.84.023813}. Field decay for gap frequencies is \(E(z) = E_0 \exp(-2z/L)\) and in the propagation region is \(E(z)=E_0\exp(-2z/L)\cos(kz)\), where $L^{-1} = L_{loc}^{-1} + L_{abs}^{-1} + L_{out}^{-1}$, with $L_{loc}$, $L_{abs}$, $L_{out}$ and $k$ being the localization length, absorption length due to inelastic scattering, scattering length due to out of plane losses, and effective propagation vector respectively. In the literature, it has been commented that absorption due to inelastic scattering does not hinder the localization process, while it reduces the overall intensity of the field profile \cite{articleYosefin} and does not provide a cutoff length analogous to an inelastic scattering length \cite{PhysRevB.47.1077}. For band edge frequencies, the presence of an electromagnetic mobility edge has been demonstrated, characterized by an unusual increase in absorption due to localization fluctuations in photon diffusivity within a disordered medium \cite{PhysRevLett.53.2169,articleYosefin}. In this analysis, it is to be noted that the absorption under consideration arises from the interaction of probe field with $\Lambda$-type atoms, which is distinct from absorption resulting from inelastic scattering processes. 

The considered PC in our analysis, has a central defect layer which is non-uniformly doped with $\Lambda$ type atoms. The structural disorder in the crystal is introduced via random variations in the refractive index of one of the bilayers. The coherent control in PC is achieved by non-uniform doping. In order to demonstrate the properties of PC, we have employed the conventional transfer matrix approach.
Our research primarily focuses on numerically investigating the coherent control of LE and absorption introduced by the defect layer containing three-level atoms in both bandgap and band edge regions for a selected range of defect layer width in the presence of disorder.  We demonstrate that both gap and band frequencies are sensitive to losses in contrast to \cite{PhysRevB.47.13120} in which it is shown that gap frequencies are independent of losses. The incorporation of atoms induces specific atomic transitions that facilitate the manipulation of effective refractive index of the crystal, thereby enabling precise control over confinement within the band gap of crystal reflected by equidistant modes. 
 
Our research uniquely reveals that absorption arising from atomic interactions yields distinct localization effects for band gap and band edge frequencies.
Notably, in the band gap region, photonic mode confinement arises as a result of changes in photonic density of states due to coherence and interference. When atomic transitions are in resonance with the probe field, atom is strongly coupled to the field due to sub-wavelength confinement. As a result, it leads to strong light localization manifested as distinct bright spots. In contrast, for off resonant conditions interference between the quantized modes weakens localization, resulting in less pronounced or diminished bright spots in absorption spectra. Further, disorder in the crystal disrupts this confinement.
Further, we highlight that at band edges of disordered PCs, light localization emerges from the interplay between multiple mechanisms rather than a single dominant effect. Structural disorder introduces Anderson-type localization through random scattering, while the coherently controlled $\Lambda$-type atomic system creates non-Hermitian regions with varying absorption. The defect layer’s high absorption (loss) and host PC’s low loss create an asymmetric potential for light. As a result, photons preferentially localize in the lower-loss regions (host PC) near the defect layer, suppressing propagation into the lossy defect layer. It enables loss-difference induced localization which is enhanced with disorder. It is to be noted that the considered PC in the absence of disorder behaves as a non-hermitian system as already reported in \cite{GHANGAS2024115703}. A recent study also shows that by creating absorption differences in a non-Hermitian honeycomb photonic lattice, Dirac points are converted into flat bands that trap light in low-loss regions, enabling tunable light control \cite{PhysRevLett.131.013802}. These mechanisms don't merely coexist but synergistically enhance each other. In our analysis, disorder increases sensitivity to loss contrasts, while non-Hermitian effects amplify the impact of structural randomness. This cooperative interaction is particularly pronounced at band edges, where the inherently slow light and enhanced density of states dramatically amplify both mechanisms. The resulting localization exhibits hybrid characteristics responding to both coherent atomic control and structural parameters. In addition, Lyapunov exponent spectra, which measure localization lengths, show a consistent pattern of coherent control across both the band gap and band edge frequency regions. This robust behavior persists even as disorder is introduced, highlighting the stability and resilience of the underlying localization mechanisms. 
Hence, we emphasize that our approach, in contrast to static disorder in \cite{Yuan:19} where the refractive index remains fixed, enables dynamic modulation of the effective refractive index, offering real-time coherent control over absorption, LE and localization.

Since PCs are predominantly confined to frequencies from visible to near-infrared (Vis-NIR) spectral range (400-1600 nm) for their applications as biosensors. In our study, we focus on Vis-NIR regime, where uncorrelated structural disorder introduces effective medium inhomogeneity. This randomness amplifies light-matter interactions through multiple scattering and localized resonances, significantly improves biosensor selectivity validated in tumor screening and blood detection.

This paper is organized as follows: In section \ref{sec2}, we introduce Lyapunov exponent and absorption using Transfer matrix method. In section \ref{sec3}, we present numerical results to demonstrate the localization dynamics with the combined effects of disorder, Lyapunov exponent and absorption. Finally, we conclude the article in section \ref{sec4}.
\section{Model}\label{sec2}
\begin{figure}[htbp]
\centering
\includegraphics[width=\linewidth]{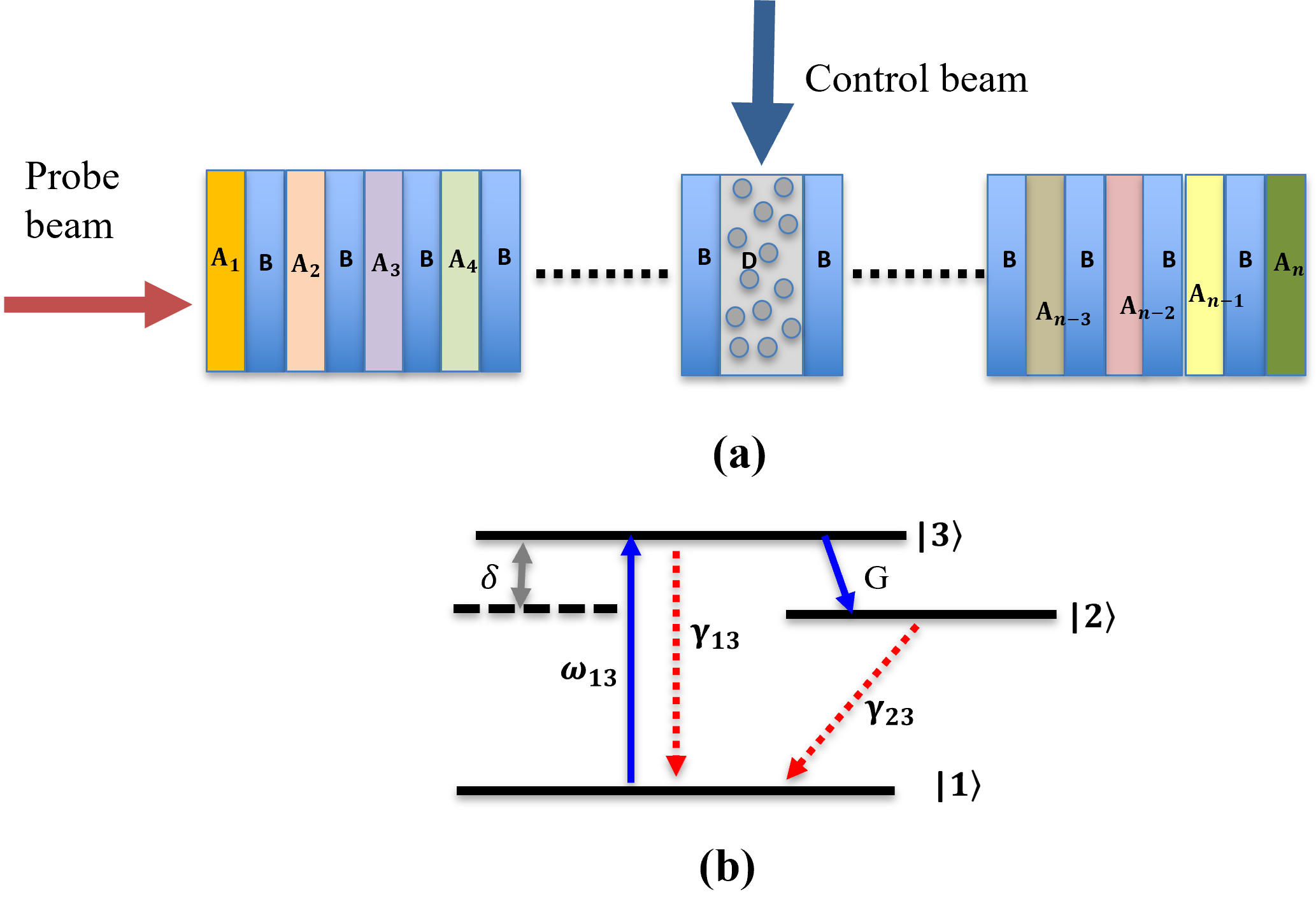}
\caption{(a) Schematic diagram of disordered photonic crystal (b) energy level configuration of $\lambda$-type atoms doped in defect layer,D. G is the rabi frequency of control field, $\gamma_{1i}$'s are the decay rates of atomic levels, $\delta$ is the detuning of probe field, $\omega$ with transition frequency of atoms $\omega_{13}$.}
\label{fig.1}
\end{figure}
A PC of length $L$ is considered as shown in Fig.\ref{fig.1} such that A and B are the alternative layers with corresponding refractive indices $n_A$, $n_B$ and $\lambda_0/4n_i, i\in A,B$, $\lambda_0=632.8 nm$ their optical thicknesses, respectively. In the middle there is a defect layer D with refractive index $n_D=2.1$ such that the configuration is of the form $(A_{1}BA_{2}BA_{3}..BDB..BA_{n-1}BA_{n})$. A control field with rabi frequency $2G$ is applied. The refractive indices of A layers are made random such that $n_{A_i} = n_{A}+\delta_j$, $\delta_j$ is the Gaussian distribution function written as $\frac{1}{2\pi\sigma}e^{-\frac{(x_{i}-\mu)^2}{2\sigma^2}}$, where $x_i = X[rand_i(0,1)]$ are random numbers and $\sigma$, $\mu$ are standard deviation and mean of $x_i$'s respectively. Random parameter $X$ (= 10, 5, 2.5, 1.25) governs the variation ranges of $\delta_j$ from $0$ to $0.05,0.12,0.20,0.42)$ where higher $X$ reduces randomness. The refractive index $n_A=2.22$ is fixed, while $n_{A_i}$ varies randomly across $A_i$ layers.
The defect layer is doped with a three-level $\Lambda$ type atomic system. The susceptibility of the doped three level atoms is of the form reported in\cite{GHANGAS2024115703}.
The transfer matrix for entire system is $M = M_{A_1} M_{B} M_{A_2} M_{B}.........M_{A_{N-1}} M_{B} M_{A_N}$
The matrix for single layer is taken as in \cite{GHANGAS2024115703}.
The transmission $T$, reflection $R$, and absorption $A$ from left to right through crystal using elements of transfer matrix are $T = M_{11}-\frac{M_{12}M_{21}}{M_{22}}$, $ R = \frac{-M_{12}}{M_{22}}$ and $A = 1-(R+T)$ respectively.
For a long enough system $n \in \infty$, the LE is given as $L(\omega)= -\frac{1}{n}\langle{\ln T}\rangle$ which is an averaged quantity over several random configurations \cite{SARABANDI2020126263}.

\section{Results and Discussions}\label{sec3}
\subsection{Effect of Control field and disorder on Lyapunov exponent and absorption}

\begin{figure}[htbp]
\centering
\includegraphics[width=\linewidth]{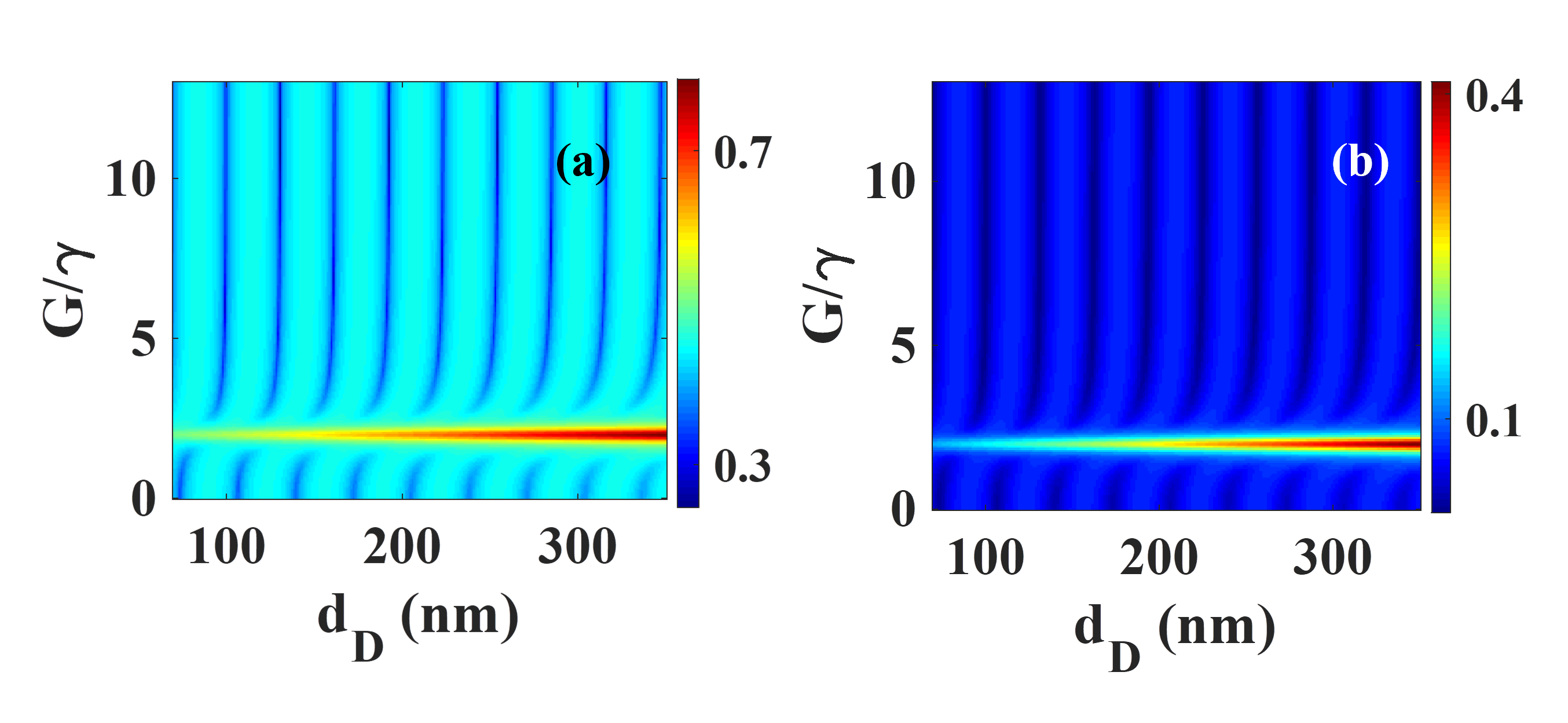}
\caption{Lyapunov exponent as a function of control field, $\frac{G}{\gamma}$ and defect layer width $d_D$ at random parameter X = 10 (a) for defect mode, $\omega = 2.3036\times 10^{15} Hz$ (b) for band edge mode, $\omega = 2.2789\times 10^{15} Hz$, for $(2n+1)$ layers, $n = 36$, $\gamma=1 MHz$.}
\label{fig.2}
\end{figure}

We numerically explore the dependence of LE on the defect layer width, $d_D$ and control field $G$ in disordered PCs. The LE inversely correlates with the penetration length, which is lower for the band gap region and decreases as the frequency approaches from edge to the band gap center. In this analysis, we have considered LE ensemble averaged over 16 random configurations. Our findings show that, for defect mode frequency, $\omega = 2.3036\times 10^{15} Hz$ lying within the band gap of disordered PC, LE exhibits maximum value ($\ge 0.7$) for a narrow range of control field [Fig.\ref{fig.2}(a)].
Unlike prior studies, our results demonstrate that the control field and defect layer properties have a pronounced effect on regions with high LE (low penetration length) within the band gap frequencies. This effect leads to enhanced surface localization, suggesting that back scattering remains dominant in this region due to destructive interference. In other words, due to minimum localization length in this region, field does not span the crystal and remains intact on the surface.
While for the region lying outside the red region, LE has comparably lower values($0.25$) such that penetration length is raised to reach enough probe field upto defect layer to induce atom-field interactions. The atomic doping of the defect layer leads to the emergence of a number of discrete states in this region. These equidistant modes observed in both the band gap and band edge regions reflect mode confinement as a result of interference. They enhance wave propagation through a mechanism akin to resonant tunneling. Hence, the possibility of field to be localized at the defect modes in gap region is increased at specific defect layer widths to induce atom field resonance. 

\begin{figure}[htbp]
\centering
\includegraphics[width=\linewidth]{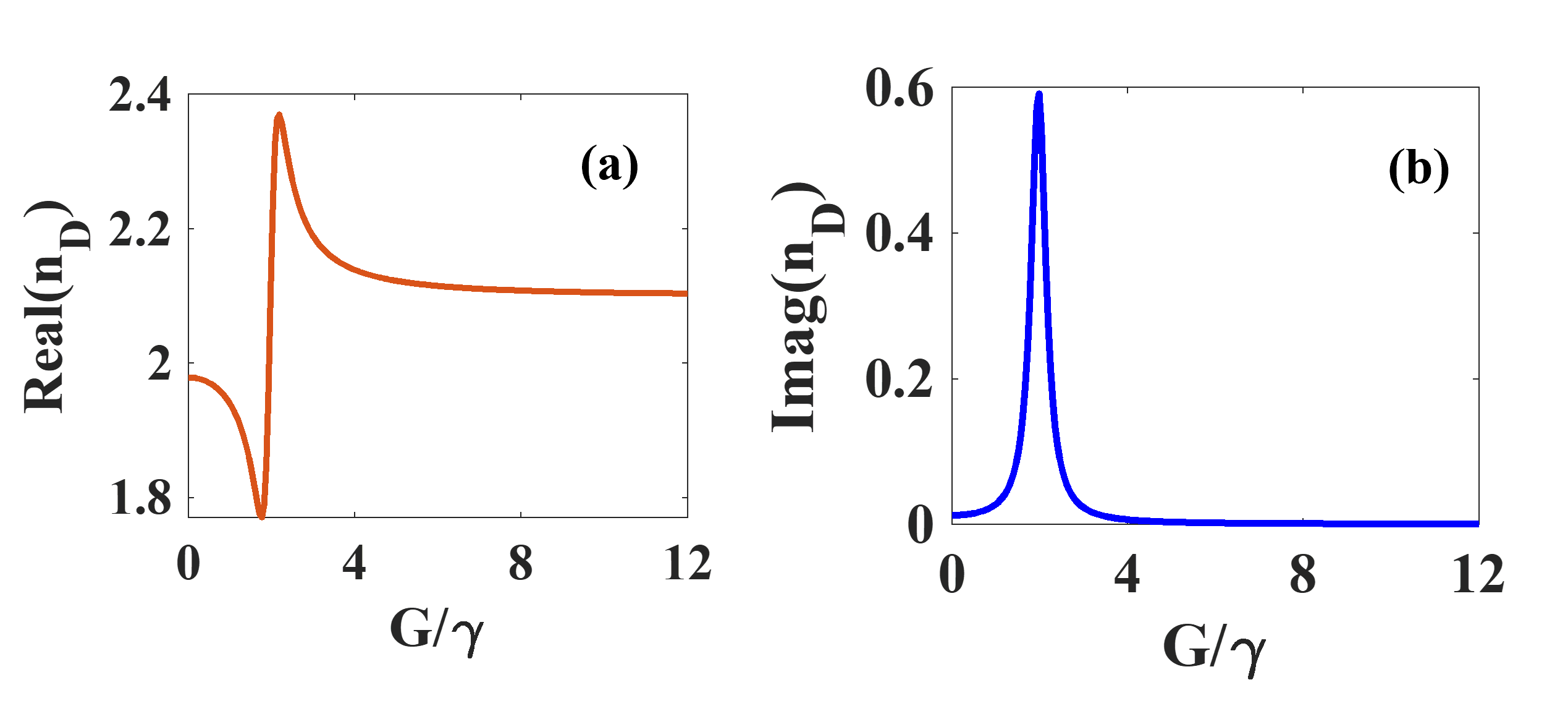}
\caption{Variation of refractive index of defect layer, $n_D$ with control field, $\frac{G}{\gamma}$ at random parameter, X = 10 (a) real component of $n_D$ (b) Imaginary component of $n_D$. Other parameters remain same as in Fig.\ref{fig.2}.}
\label{fig.3}
\end{figure}
\begin{figure}[htbp]
\centering
\includegraphics[width=\textwidth]{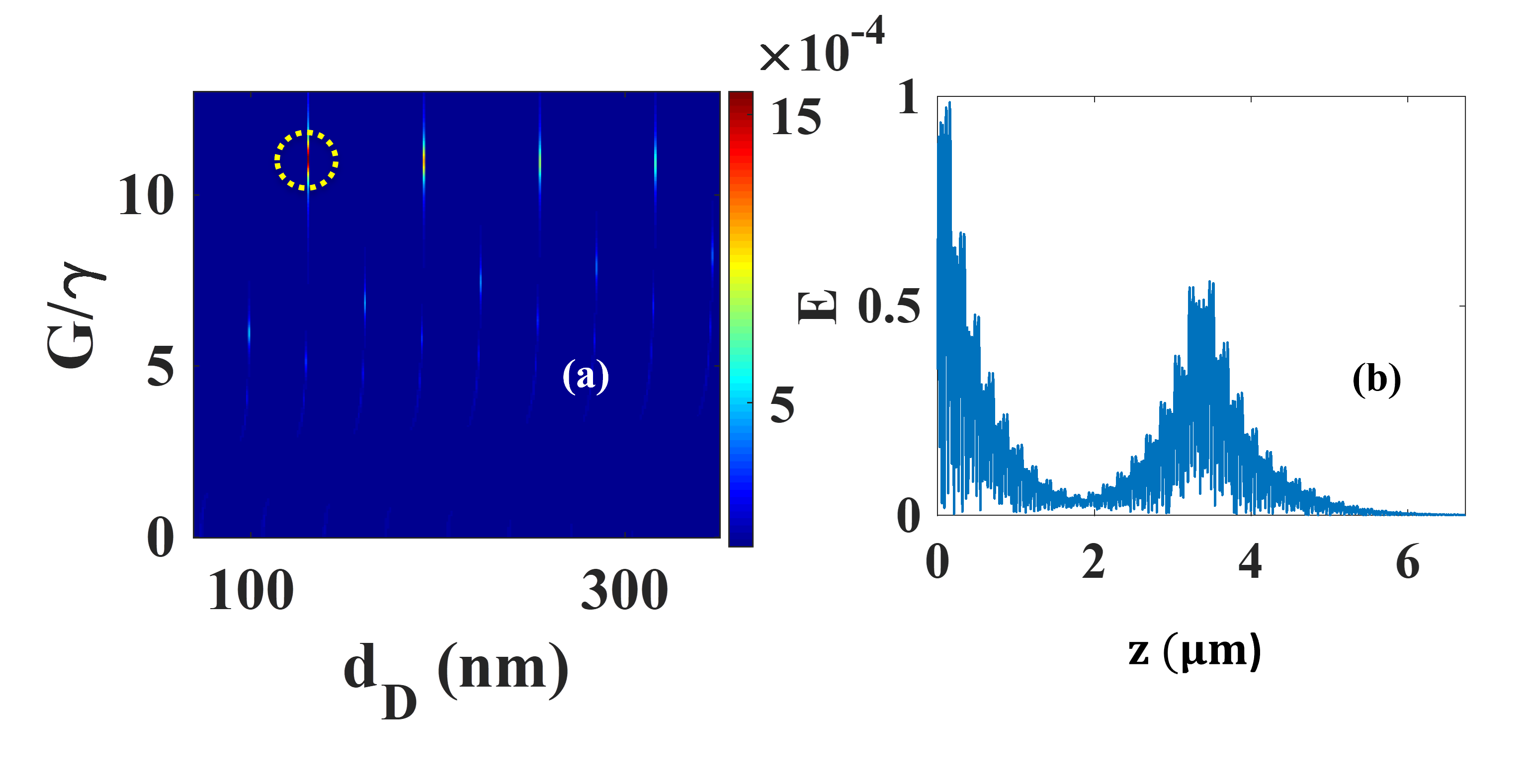}
\caption{(a) Absorption spectra as a function of control field, $\frac{G}{\gamma}$ and defect layer width $d_D$ for defect mode for randomness parameter X = 10 at $\omega=2.3036 \times 10^{15} Hz$, (b) field profile along crystal axis, z for region marked in (a) in yellow ring at $G = 11.0683\gamma$, $d_D = 130.7229 nm$, for $(2n+1)$ layers, n = 36.}
\label{fig.4}
\end{figure}
\begin{figure}[htbp]
\centering
\includegraphics[width=\textwidth]{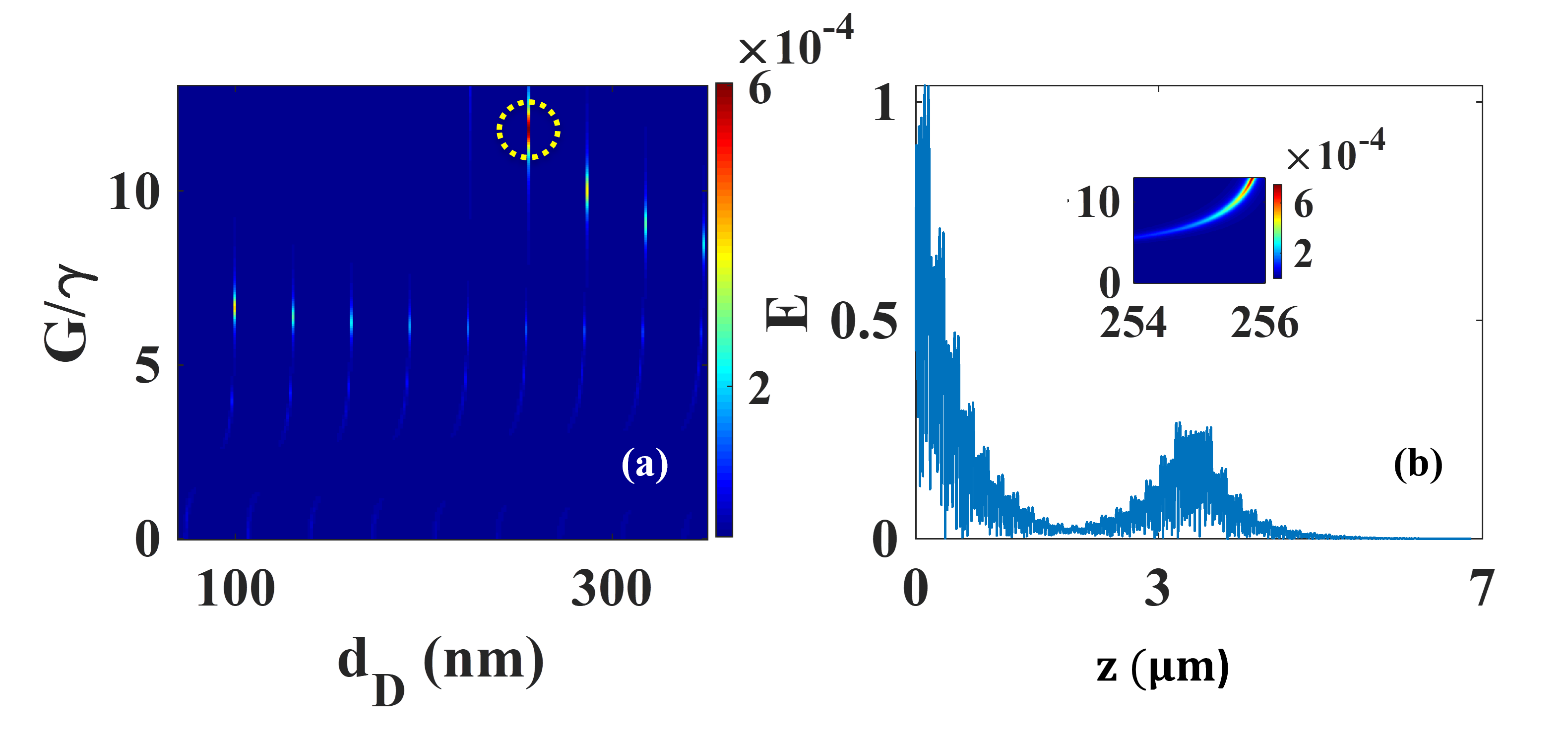}
\caption{(a) Absorption spectra as a function of control field, $\frac{G}{\gamma}$ and defect layer width $d_D$ for defect mode for randomness parameter X = 5 at $\omega=2.2874 \times 10^{15} Hz$ (b) field profile along crystal axis, z for region marked in (a) in yellow ring at $G = 11.7588\gamma$, $d_D = 255.7286 nm$, for $(2n+1)$ layers, n = 36.}
\label{fig.5}
\end{figure}
\begin{figure}[htbp]
\centering
\includegraphics[width=\textwidth]{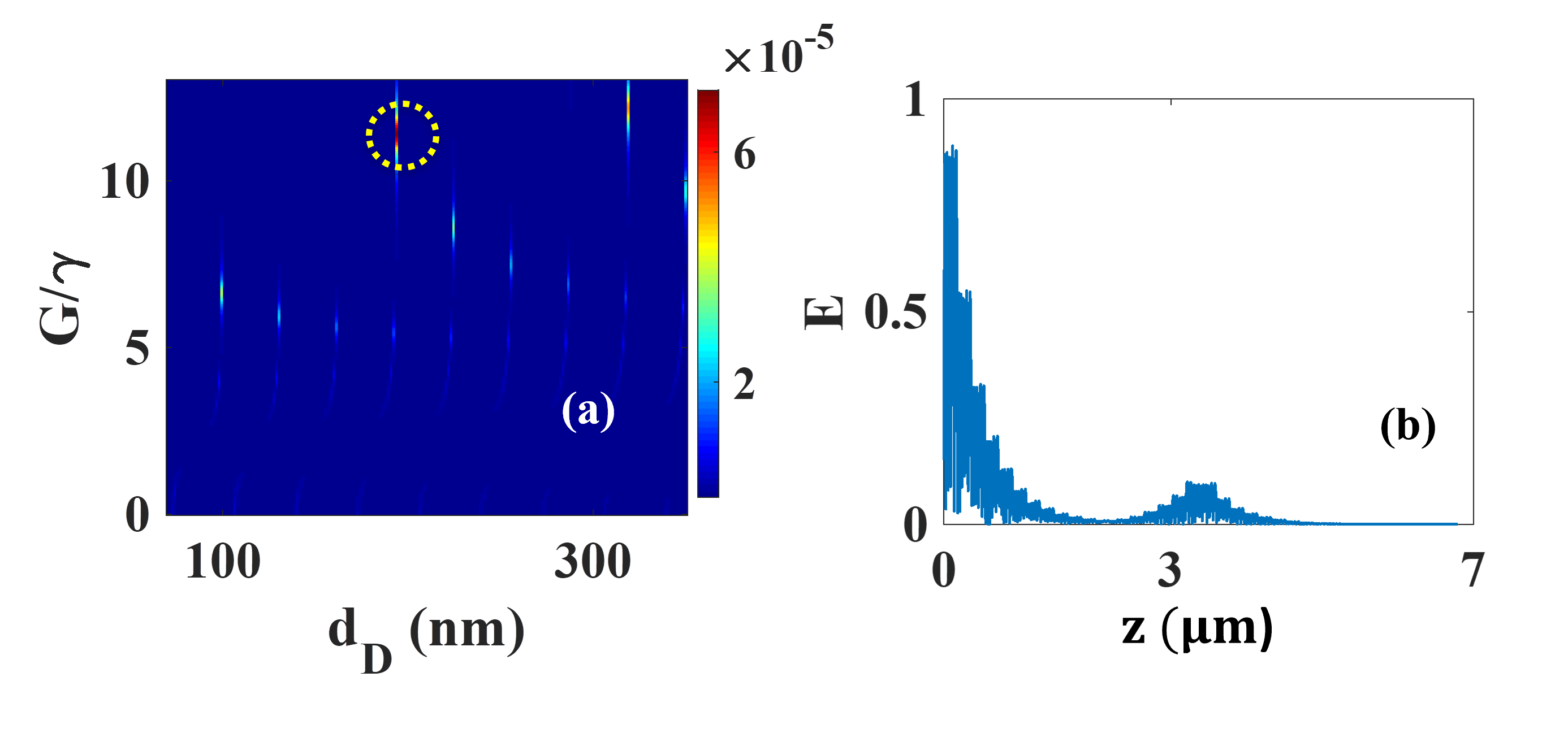}
\caption{(a) Absorption spectra as a function of control field, $\frac{G}{\gamma}$ and defect layer width $d_D$ for defect mode for different randomness parameter X = 2.5 at $\omega=2.2772 \times 10^{15} Hz$ (b) field profile along crystal axis, z for region marked in (a) in yellow ring at $G = 11.4322\gamma$, $d_D = 193.8191 nm$, for $(2n+1)$ layers, n = 36.}
\label{fig.6}
\end{figure}
For band edge frequencies, LE exhibits a similar trend as for gap region but its maximum amplitude is smaller ($\le 0.4$) as compared to gap region ($\le 0.75$) due to the presence of higher DOS [Fig.\ref{fig.2}(b)]. It means that most of the probe field will be propagated through crystal but for narrow region of control field, $G=0.15\gamma$, backscattering remains dominant as for the case of defect mode frequency [Fig.\ref{fig.2}(a)]. It has already been demonstrated that in disordered PC cavities, Anderson localization causes the DOS to saturate beyond a critical length, indicating that local disorder, rather than cavity boundaries, governs light confinement. It can be calculated by Bloch mode expansion using Green's function in 3D systems \cite{doi:10.1021/acsphotonics.7b00967}. Previous studies have shown that DOS in finite 1D PCs exhibits distinct band gap and band edge features \cite{PhysRevB.89.045123}. Further, tailoring the DOS can control Anderson localization in PC waveguides, with disorder having minimal impact on these spectral characteristics \cite{PhysRevB.82.165103,PhysRevB.107.014205}.

Furthermore, the variation of real and imaginary components of refractive index of defect layer are shown in [Fig.\ref{fig.3}]. It is observed that corresponding to narrow red region in LE spectra, real and  imaginary component  of refractive index of defect layer shows maximum variation [Fig.\ref{fig.3}]. While for other G values refractive index remains saturated. These variations lead to the change in effective refractive index of the PC and alters its overall transmission. 
\begin{figure}[htbp]
\centering
\includegraphics[width=\textwidth]{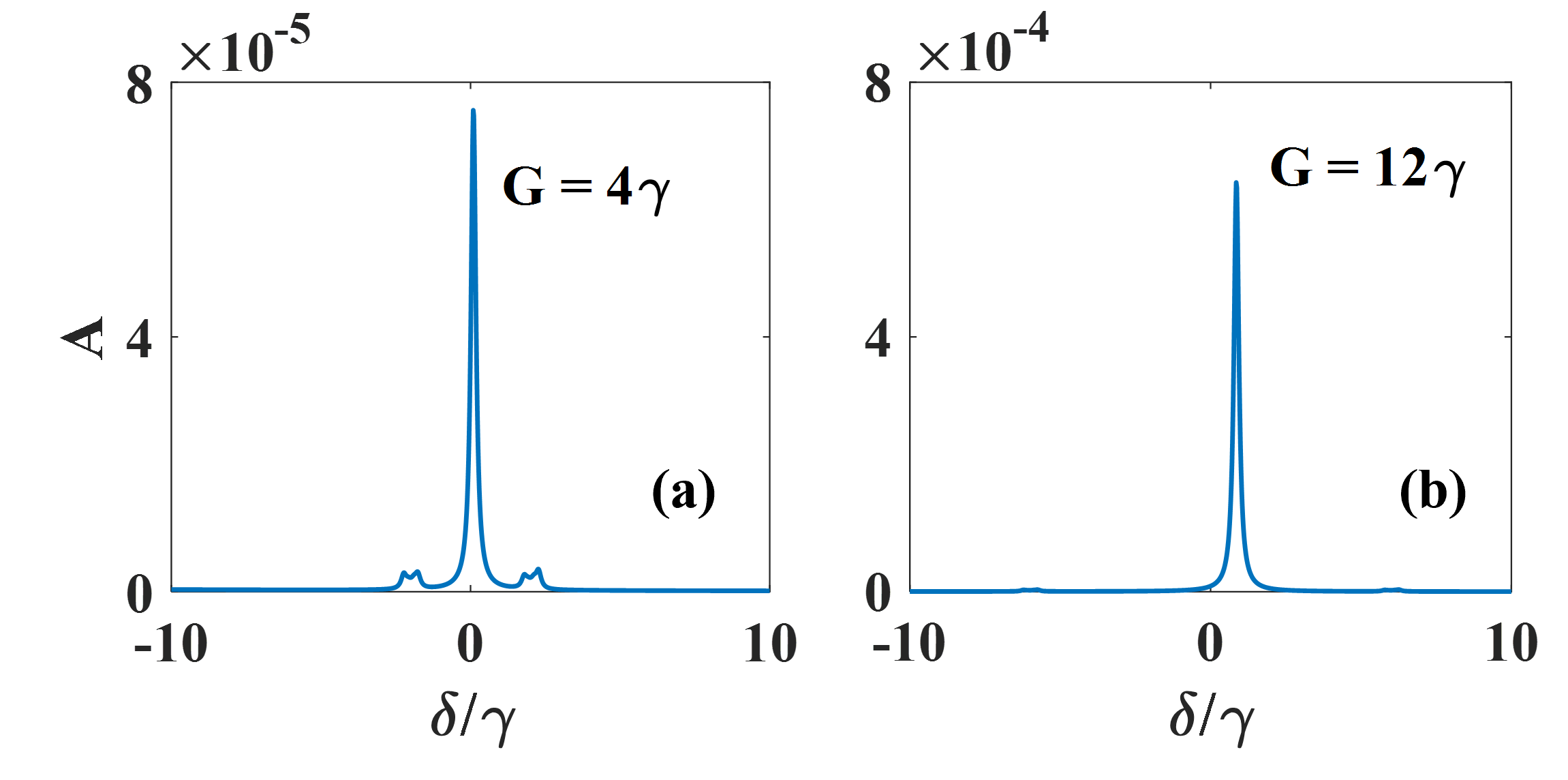}
\caption{(a) Absorption as a function of detuning, $\frac{\delta}{\gamma}$, for randomness parameter $X = 5$ at $\omega=2.2874 \times 10^{15} Hz$ and defect layer width $d_D = 255.7286 nm $ at different control fields (a) $G=4\gamma$ (b) $G=12\gamma$, for $(2n+1)$ layers, n = 36.}
\label{fig.7}
\end{figure}

\begin{figure}[htbp]
\centering
\includegraphics[width=\textwidth]{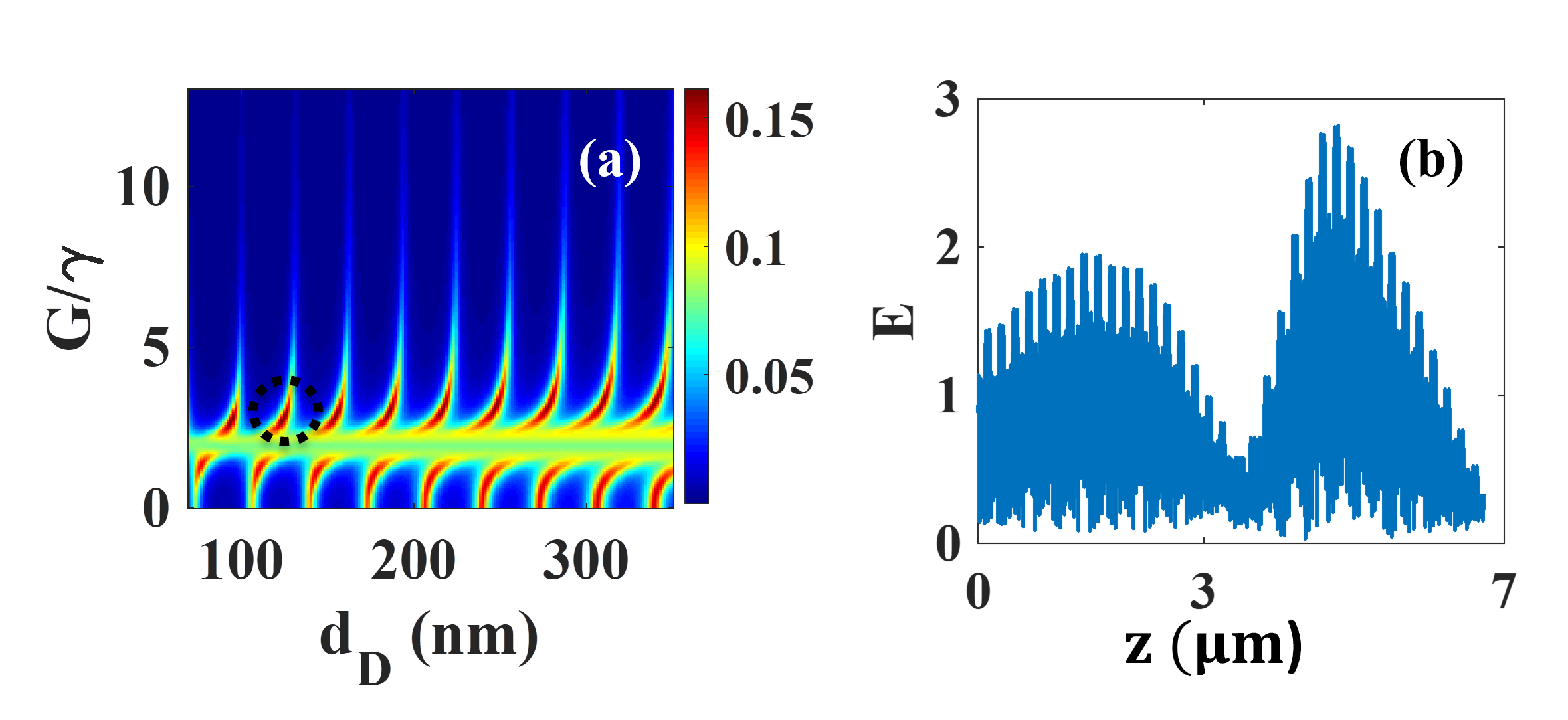}
\caption{(a) Absorption spectra for band edge mode as a function of control field, $\frac{G}{\gamma}$ and defect layer width $d_D$ for  random parameter, X = 10 at $\omega=2.2789 \times 10^{15} Hz$ (b) electric field profile along crystal axis, z at $G = 2.809\gamma$, $d_D = 124.8744 nm$ marked in black ring in (a), for layers $2n+1$, n = 36.}
\label{fig.8}
\end{figure}

\begin{figure}[htbp]
\centering
\includegraphics[width=\textwidth]{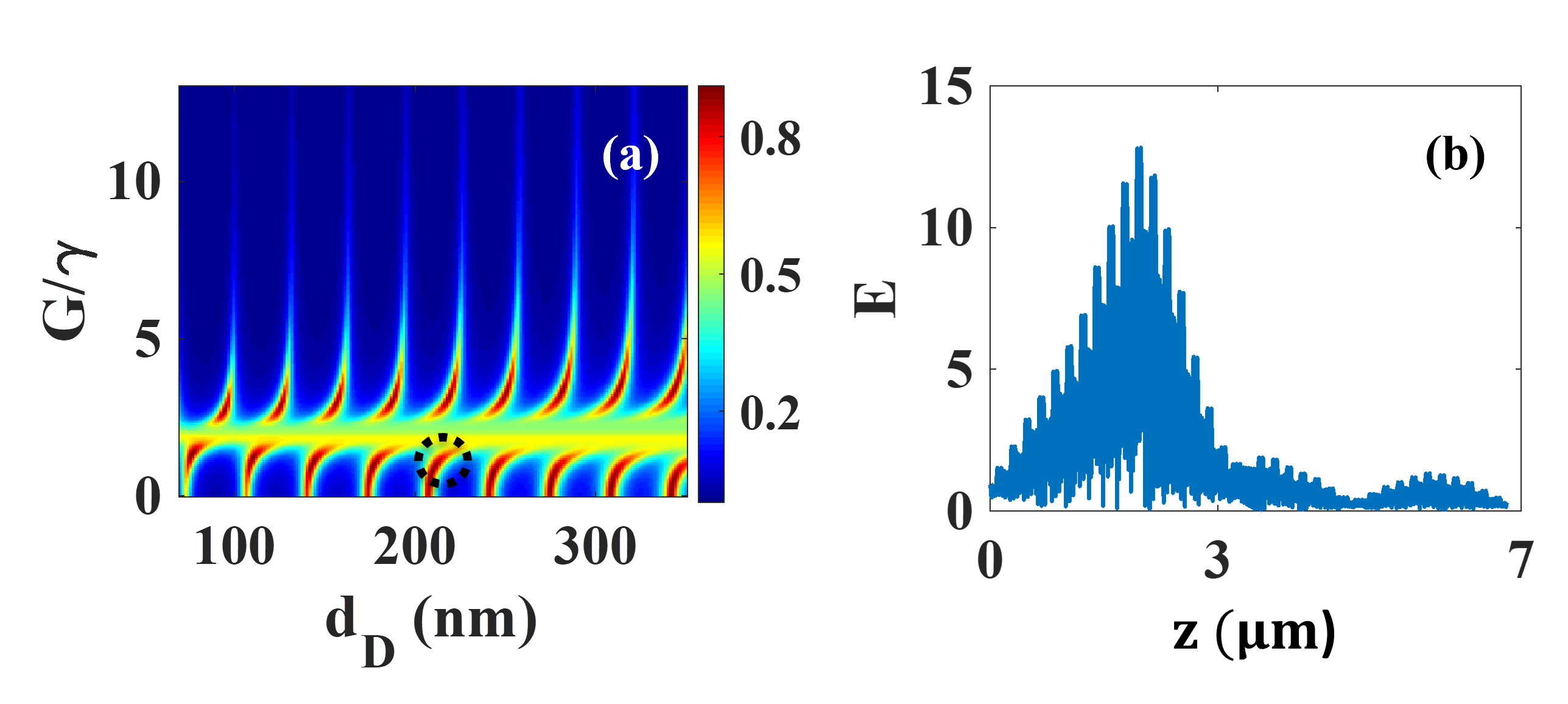}
\caption{(a) Absorption spectra for band edge mode as a function of control field, $\frac{G}{\gamma}$ and defect layer width $d_D$ for  random parameter X = 5 at $\omega=2.2525 \times 10^{15} Hz$ (b) electric field profile along crystal axis, z at $G = 0.71859\gamma$, $d_D = 209.2965 nm$ marked in black ring in (a), for layers $2n+1$, n = 36 }
\label{fig.9}
\end{figure}

\begin{figure}[htbp]
\centering
\includegraphics[width=\textwidth]{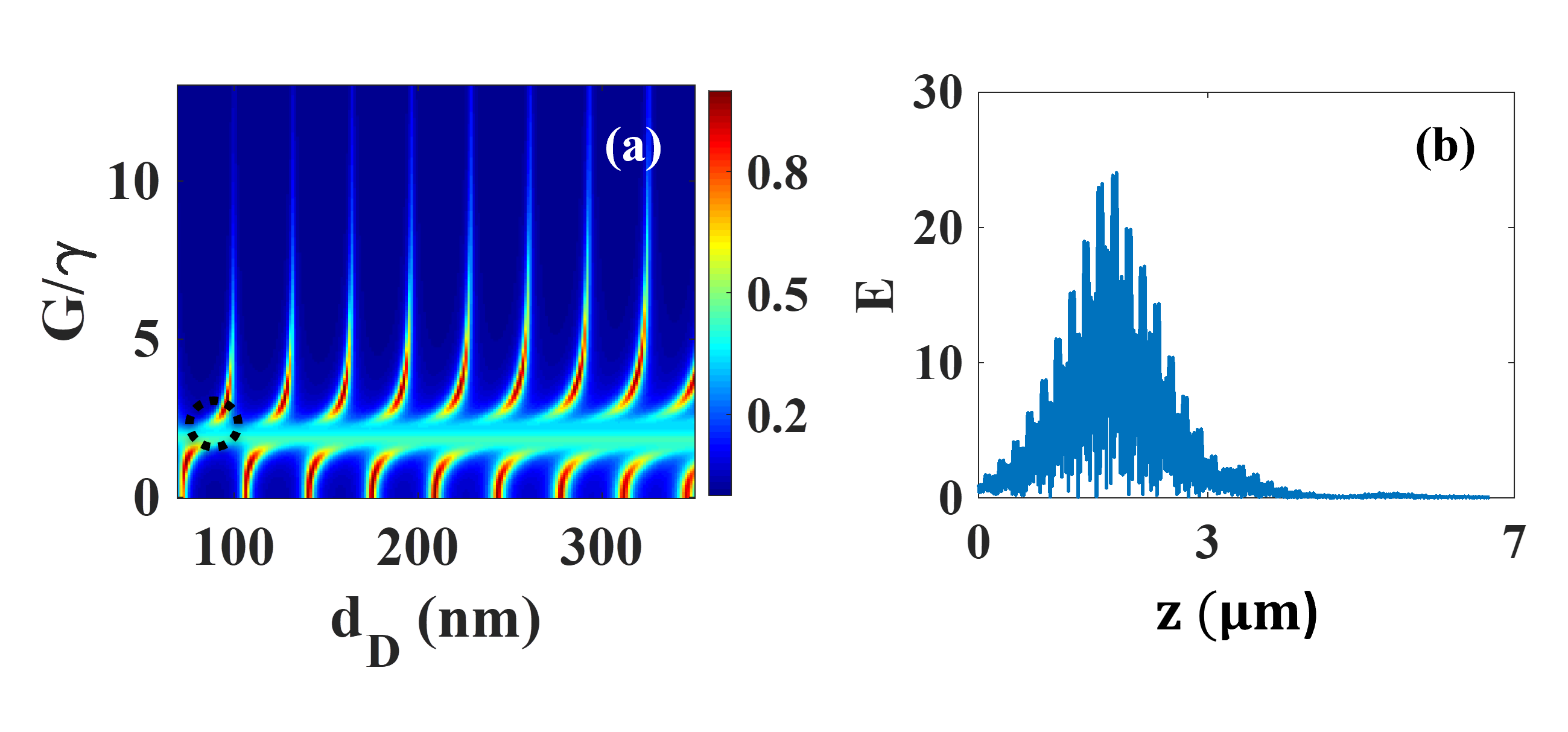}
\caption{(a) Absorption spectra for band edge mode as a function of control field, $\frac{G}{\gamma}$ and defect layer width $d_D$ for randomness parameter, X = 2.5 at $\omega=2.2131 \times 10^{15} Hz$, (b) electric field profile along crystal axis, z at $G = 1.1106\gamma$, $d_D = 74.2211 nm$ marked in black ring in (a), for layers $2n+1$, n = 36 }
\label{fig.10}
\end{figure}

Hence, the coherent control in LE spectra is attributed to the modifications in the refractive index of the doped defect layer and consequently of the PC. It leads to the formation of quantized states in LE spectra and also observed in absorption spectra discussed later. The band gap in PCs is usually accompanied by smaller absorption because most of the field is backscattered due to destructive interference caused by crystal's periodicity. We observe that because of the presence of doped defect layer, phenomenon of optical mode confinement takes place and a number of discrete defect modes arise as the control field modifies the refractive index of the defect layer. These modes are accompanied with the field confinement and small spots of highly absorbed trapping light (marked in yellow ring in [Figs.\ref{fig.4},\ref{fig.5},\ref{fig.6}]). These highlighted absorption hot spots indicate the resonance of doped atoms with the probe field in the disordered cavity. It significantly increases local field intensity at the site. Since light is confined and cannot escape due to bandgap, localized field interacts more efficiently with the absorbing medium, reinforcing the localization effect. The absorption is minimal in regions with high LE values because the field is destructively quantized due to minimum localization length and is not sufficient to interact with atoms. Inset in Fig.\ref{fig.5}(b) displays high-resolution absorption plot corresponding to the atomic resonance with high Q defect mode reflecting very small mode volume. In contrast, the behaviour of absorption reverses for band edge frequencies corresponding to the narrow range of control field (for which LE is maximum) as shown in Figs.\ref{fig.8},\ref{fig.9},\ref{fig.10}. Absorption gets maximum at the band edges corresponding to region with maximum LE value ($~0.4$) [Fig.\ref{fig.2}(b)]. It might happen due to enhanced DOS at band edges. On the other hand, absorption is minimum in other regions where the overlap between the atomic transitions and band edge modes is reduced. This might happen due to saturation effects at high DOS, where the atomic transitions become saturated, leading to reduced probe field interaction despite high DOS. This coherent control of absorption at band edges enables the manipulation of electromagnetic mobility edge accompanied with high absorption and refers to a transition point separating localized electromagnetic modes (immobile) from extended modes (mobile) in a disordered or periodic medium \cite{PhysRevLett.53.2169}.
Additionally, we observe the strong coupling behaviour at these absorption bright spots lying in band gap regions. We report that absorption variation with detuning at low control field, $G=4\gamma$ yields multi-peak absorption spectrum as shown in Fig.[\ref{fig.7}(a)]. These multi modes arise in band gap due to the combined effect of coherence and randomness. The side peaks resemble Autler–Townes splitting due to associated anti-crossing behavior. The strong coupling between the defect mode and collective atomic transitions prevents level crossing and increases side peaks' separation with increase in control field strength. Meanwhile, the sharpening of the central absorption peak at $G=12\gamma$ is accompanied with increase in absorption amplitude (to the order of $10^{-4}$ from $10^{-5}$) due to interference between a strong control field and probe field. It signals resonance of defect mode with collective atomic transitions, resulting in tighter light confinement at that frequency as shown in Fig.[\ref{fig.7}(b)]. In contrast Autler-Townes splitting at higher field strength have very low (almost vanishing) absorption amplitudes. Hence, we limit our discussion to these key findings. However, there remains significant potential to explore quantum-scale interactions between atomic ensembles and optical defect modes. Recent work has shown that by engineering coupled single-mode cavities, one can achieve tunable dark and bright modes via perfect destructive interference, classical interference arises from entangled bright and dark quantum states, and multilevel atoms in optical cavities can form entangled dark states through collective dissipation \cite{WANG2019106,villasboas2024brightdarkstateslight,PhysRevX.12.011054}.

\begin{figure}[htbp]
\centering
\includegraphics[width=\textwidth]{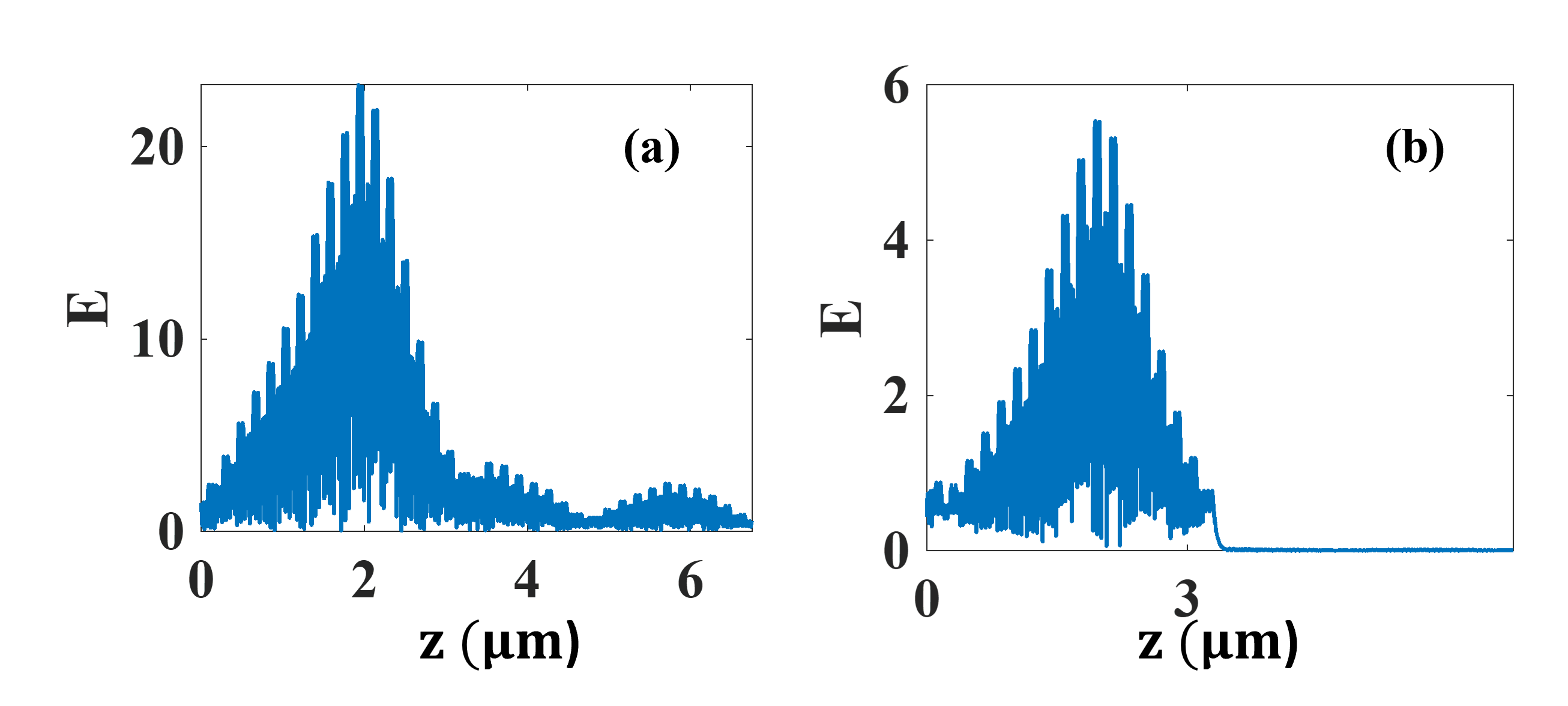}
\caption{Electric field profiles along crystal axis, z from Fig.\ref{fig.7}(a) for different regions,(a) Yellow region, at $G = 1.9\gamma$, $d_D = 150nm$ (b) Blue region, at $G = 6\gamma, d_D = 150nm$, parameters are same as in Fig.\ref{fig.7}.}
\label{fig.11}
\end{figure}
\subsection{Variation of field profile}
We further investigate the combined effects of randomness, LE, and absorption on the localization properties of disordered crystal with $2n+1$ layers, $n = 36$. We numerically examine electric field profiles corresponding to absorption and LE to obtain insights on localization across the length of disordered PC.
We know for a crystal system, 
We observe that within the band gap region (defect modes) field amplitude is significantly confined within the defect layer corresponding to marked hot spots of maximum absorption due to confinement localization [Figs.\ref{fig.4},\ref{fig.5},\ref{fig.6}(a)]. When atomic transitions are in resonance with the probe field, atom is strongly coupled to the field due to sub-wavelength confinement. As a result, it leads to strong light localization manifested as distinct bright spots. In contrast, for off resonant conditions interference between the quantized modes weakens localization, resulting in less pronounced or diminished bright spots in absorption spectra. However, as randomness increases from $X=10$ (low disorder) to $X=2.5$ (high disorder), it leads to the suppression of confined field amplitude in defect layer from $0.55$ to nearly $0.1$ [Figs.\ref{fig.4},\ref{fig.5},\ref{fig.6}(b)].

In contrast, at band edge frequencies, the interplay of structural randomness and atomic loss difference enhances localization near the defect layer, a region characterized by low-loss [Figs.\ref{fig.8}, \ref{fig.9}, \ref{fig.10}]. For PCs with low disorder $X=10$ [Fig.\ref{fig.8}(b)], the field profile is localized and exhibit uneven extensions in low-loss regions. As disorder increases to $X=5$, the oscillatory field profile is progressively dampened and becomes strongly localized in the vicinity of defect layer [Fig.\ref{fig.9}(b)] and $X=2.5$ [Fig.\ref{fig.10}(b)]. This transition culminates in the emergence of Anderson localization with loss difference induced localization where disorder increases sensitivity to loss contrasts, and non-Hermitian effects amplify the impact of structural randomness. These two effects synergistically lead to the evolution of localization from weak to strong with increasing disorder in PC.
Further, we can also relate this type of localization with critical states ($z^{-L}$) that appeared in bandgap region with nonlinearity as control parameter as discussed in \cite{PhysRevB.41.8047} but in that case system is made of non-linear Fibonacci multilayers.
In contrast to the bandgap behavior, high absorption at band edge frequencies suppresses the field amplitude within the defect layer, consistent with observations reported in \cite{articleYosefin,PhysRevB.47.1077}.

Additionally, we analyze the field profiles at different control field and defect layer width values at band edge modes. In the high-absorption region (yellow region in Fig.\ref{fig.9}(a)) at $G=6\gamma$ and $d_D=150 nm$, the field intensity rapidly decays upon reaching the defect layer, likely due to the combined effects of high LE and absorption [Fig.\ref{fig.11}(b)]. As we transition toward regions with lower absorption in Fig.\ref{fig.9}(a)) at $G=1.9\gamma$ and $d_D=150 nm$, the field profile becomes oscillatory though damped and critical states arise [Fig.\ref{fig.11}(a)]. 

\begin{figure}[htbp]
\centering
\includegraphics[width=\textwidth]{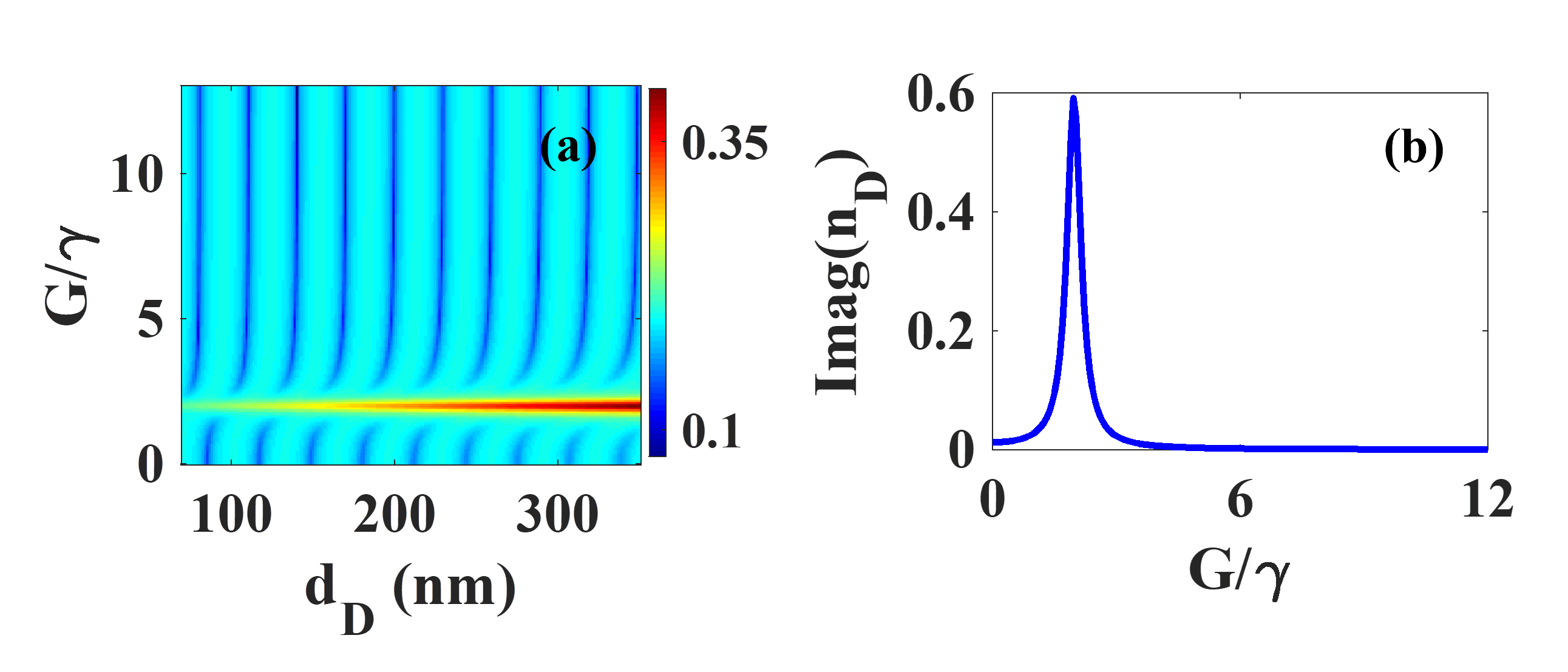}
\caption{(a) Lyapunov exponent as a function of control field, G and defect layer width, $d_D$, (b) Imaginary component of refractive index of defect layer, $n_D$ at $\omega=2.3936 \times 10^{15} Hz$, for $2n+1$ number of layers, n = 60 at random parameter, X = 5.}
\label{fig.12}
\end{figure}

\begin{figure}[htbp]
\centering
\includegraphics[width=\textwidth]{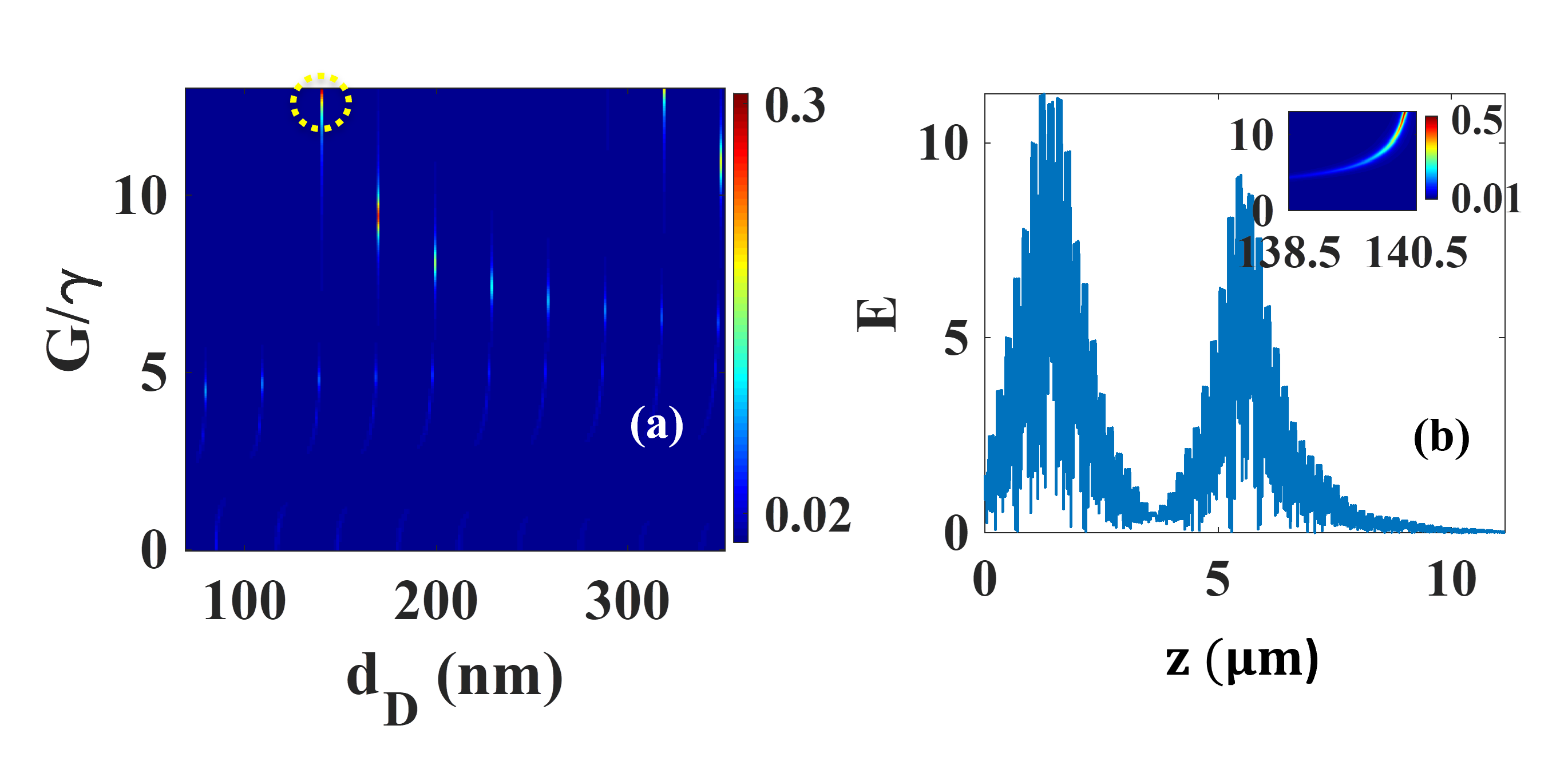}
\caption{(a) absorption spectra at $\omega=2.3936 \times 10^{15} Hz$ as a function of control field, G and defect layer width, $d_D$ and (b) field profile along crystal axis, z at $G=13\gamma$, $d_D = 140.3518 nm$, inset in (b) depicts resonant mode highlighted in yellow ring in (a), for $2n+1$ number of layers, n = 60 at random parameter, X = 5.}
\label{fig.13}
\end{figure}

We have also analyzed the possibility of localization due to disorder and control field across the band gap region for crystal with n = 60. In Figs.\ref{fig.12} and \ref{fig.13}, the behaviour of LE, refractive index of defect layer, absorption, and electric field profile respectively is depicted at different gap frequency $\omega=2.3936 \times 10^{15} Hz$ lying towards right band edge. The field amplitude at the defect layer is enhanced due to the combined effect of absorption and LE at $G=13\gamma$, $d_D = 140.3518 nm$ [Fig.\ref{fig.13}(b)]. Inset in Fig.\ref{fig.13}(b) display high-resolution absorption plot corresponding to the red spot in Fig.\ref{fig.13}(a). This inset plot indicates that the resonance spots are exceedingly delicate with low mode volume confinement, a result of PC cavity's high Q factor. While the behaviour of defect modes from mid gap remains same as discussed above for n = 36 in [Figs.\ref{fig.4},\ref{fig.5},\ref{fig.6}]. We want to emphasize that due to atomic doping, the occurrence of resonant modes is possible for the entire range of gap frequencies as a result of the control field even in the presence of disorder. This trend is observed for n=20 also but results are not shown. 
Hence, this study provides valuable insights into the switching of PC properties from strong to weak localization due to disorder. Moreover, it also supports the notion of coherent control of the mobility edge and localization across disordered PC in the considered range of randomness.
Furthermore, with penetration depths ranging from approximately 7 $\mu m$ (for n=36) to nearly 11 $\mu m$ (for n=60) [Fig.\ref{fig.13}(b)], this study opens up promising applications in biosensing focusing on sample surface interactions. 

\section{Conclusion}\label{sec4}
In conclusion, our work demonstrates that the interplay between absorption, disorder, and coherent control in photonic crystals leads to two distinct localization regimes. It is accompanied with the contrasting behaviour of absorption in band gap and band edge regions.  In the band gap, strong coupling between resonant atoms and optical modes enhances absorption-driven confinement in the defect layer, although increasing disorder can disrupt this confinement. Conversely, at band edges, the synergy between disorder-induced Anderson localization and loss-difference-induced trapping strengthens field confinement near the defect layer. Furthermore, our analysis of the LE spectra which quantitatively measure localization lengths confirms that these dual localization behaviors are robust across varying levels of disorder. By non-uniformly doping the defect layer with $\lambda$-type atoms, we achieve dynamic modulation of the effective refractive index, enabling precise tuning of the mobility edge and allowing for a controlled switch between strong and weak localization states. These findings offer valuable insights for designing adaptive photonic devices with tailored optical properties, and they pave the way for future research into alternative doping geometries and multi-layered disorder architectures.









 \bibliographystyle{elsarticle-num} 
 \bibliography{references}

\end{document}